# Lenslet Light Field Image Coding: Classifying, Reviewing and Evaluating

Catarina Brites, *Member*, *IEEE*, João Ascenso, *Senior Member*, *IEEE*, and Fernando Pereira, *Fellow*, *IEEE*

*Abstract*— In recent years, visual sensors have been quickly improving, notably targeting richer acquisitions of the light present in a visual scene. In this context, the so-called *lenslet light field (LLF) cameras* are able to go beyond the conventional 2D visual acquisition models, by enriching the visual representation with directional light measures for each pixel position. LLF imaging is associated to large amounts of data, thus critically demanding efficient coding solutions in order applications involving transmission and storage may be deployed. For this reason, considerable research efforts have been invested in recent years in developing increasingly efficient LLF imaging coding (LLFIC) solutions. In this context, the main objective of this paper is to review and evaluate some of the most relevant LLFIC solutions in the literature, guided by a novel classification taxonomy, which allows better organizing this field. In this way, more solid conclusions can be drawn about the current LLFIC *status quo*, thus allowing to better drive future research and standardization developments in this technical area.

*Index Terms*—Light field image coding, lenslet, taxonomy, perspective image, micro-image.

## I. INTRODUCTION

In recent years, significant developments in visual representation technology have occurred, aiming to increase the user quality of experience (QoE) by providing highly immersive and fully realistic 3D experiences. As it is well-known, the so-called *plenoptic function* describes the intensity of light at any point in space ($x,y,z$), coming from any angular direction ($\theta, \varphi$), over time ($t$), and for each wavelength ($\lambda$) [1][2]. This means that highly immersive experiences may be provided to the users if the plenoptic function information is effectively captured and replicated. However, sensors in conventional cameras just capture the total light intensity hitting each pixel position and, thus, directional information about the light rays is lost. This is clearly a limited representation of the real scene. The recent emergence of sensors with the capability to capture higher dimensional visual representations has allowed improving the conventional imaging representation model based on 2D planes and increased the potential to offer the users high quality experiences in terms of immersion and realism. For instance, by placing a micro-lens (ML), i.e. *lenslet*, array in the optical path of a conventional monocular camera, it is possible to capture the light for each spatial position ($x, y$) and coming from any angular direction ($\theta, \varphi$). This imaging representation model, which can be seen as richer way of sampling the plenoptic function information regarding the conventional model, is known as *lenslet light field* (LLF) imaging; for length reasons, this paper will focus on the LLF coding technology corresponding to the visual information for a single time instant, i.e. a *LLF image*.

While LLF imaging is an important step forward to provide increased immersive and realistic 3D experiences, its acquisition process results in a large amount of data, which requires a significant storage space and transmission bandwidth, if efficient coding solutions are not used. For this reason, tens of LLF image coding (LLFIC) solutions have been proposed in recent years. However, a comparison of their performances is still a rather difficult task since the reported performance results have been obtained most of the times under different test conditions and evaluation methodologies and there is no public software available to obtain comparable results. Acknowledging the practical importance of developing efficient LLF coding solutions, JPEG has launched in 2015 the JPEG Pleno standardization activity, addressing light field (LF) imaging, acquired by both LLF cameras and high density camera arrays [3]. In January 2017, JPEG issued a Call for Proposals (CfP) on LF coding technologies [4], asking for efficient coding solutions fulfilling an identified set of requirements. The main goal is to standardize LF coding solutions providing interoperability between different products and applications. In JPEG, light field coding solutions will be specified in JPEG Pleno Part 2, named Light Field Coding [5], one of the parts of the JPEG Pleno standard that JPEG is planning to specify coding solutions for plenoptic imaging modalities, where light fields are considered along with point clouds and holographic data.

In this context, this paper first proposes a meaningful classification taxonomy for LLFIC solutions that allows to identify and abstract their differences, commonalities and relationships. Guided by this classification taxonomy, some of the most relevant LLFIC solutions available in the literature are then reviewed and their compression performances analyzed under precise and meaningful test conditions. It is important to stress that the main purpose of this paper is not to propose a novel LLFIC solution but rather to organize, classify and evaluate a technical area that has received many contributions in recent years. This type of paper is essential to gather a systematic, high-level and more abstract view of the field to further launch solid and consistent advancements in this technical area. With this purpose in mind, the rest of this paper is organized as follows: Section II will briefly review the LLF imaging basics, while Section III will propose a classification taxonomy for the many LLFIC solutions in the literature. Section IV will review the most relevant LLFIC solutions in the literature driven by the proposed taxonomy and, finally, Section V will present a comparative performance analysis of the LLFIC solutions reviewed in Section IV.

## II. LLF IMAGING: A BRIEF REVIEW

LLF imaging is a 3D visual representation model where the scene's light radiance is captured through a high-density set of

The authors are with Instituto Superior Técnico, Universidade de Lisboa - Instituto de Telecomunicações, Av. Rovisco Pais, 1040-001 Lisboa, Portugal (e-mails: catarina.brites@lx.it.pt, joao.ascenso@lx.it.pt, fp@lx.it.pt)

tiny lenses located in a single camera, the so-called *lenslet light field cameras*. These cameras acquire a 2D array of so-called *micro-images* (*MI*), captured using an array of MLs placed in front of the camera's photosensor. As the ML array (MLA) is placed in the optical path of a conventional monocular camera, the LLF image provides directional information for each sampled position [6]; this directional information is precisely the main added value of this type of sensor/imaging, as provides a richer visual representation.

There are two main LLF camera architectures depending on the MLA placement: the so-called *unfocused* and *focused* cameras, aka plenoptic camera 1.0 and 2.0, respectively. In the unfocused cameras, such as in the Lytro cameras' family [7], the MLA is placed at exactly one focal length in front of the photosensor plane, with MLs focused on infinity; on the other hand, in the focused cameras, such as the Raytrix cameras' family [8], the MLA is placed at a distance *b* in front of the photosensor plane, with the MLs focused on the main lens image plane. As a consequence, the information contained within and between MIs differs for each of these camera architectures. In the unfocused cameras, each MI captures only angular information (all directions) for a given spatial sample and the information regarding the spatial samples (spatial information) for a given direction are spread across MIs; on the other hand, in the focused cameras, both spatial and angular information are captured in each MI and across MIs. Besides the optical setup, these two LLF camera architectures also typically differ on the MLA structure, i.e. number of MIs as well as their shape and size.

Ideally, each ML should cover the largest number of photosensor pixels to create a MI with the highest angular resolution [9]. However, by doing this, the number of MIs would be reduced for the same photosensor resolution, and so also the spatial resolution for each angular direction. This highlights that there is an important trade-off between spatial and angular resolution in this type of cameras. While the unfocused camera favors a spatial resolution reduction as a trade-off for higher angular resolution, the focused camera works the other way around, thus favoring a spatial resolution increase at the price of a lower angular resolution [9]. The LLF images obtained directly from the sensor, so-called *raw LLF images,* need to be processed such that some extracted/rendered information can be displayed, for example, in conventional 2D or autostereoscopic displays, as native light displays are not yet available. The data rendered from the raw LLF image are known in the literature as *perspective images* (PIs) or *sub-aperture images* (SAIs), where each PI (or SAI) represents a different perspective view (or viewpoint) to the scene. Naturally, different processing and rendering techniques are used for unfocused and focused LLF content due to the different optical acquisition setups.

Due to the key role that unfocused LLF content, acquired with a Lytro Illum LF camera, will assume in sections IV and V, it is worth to briefly review here the approach adopted by the JPEG Pleno LF coding CfP [4] and Common Test Conditions (CTC) [10] to render PIs from the Lytro Illum raw LLF image. The Lytro Illum raw LLF content processing includes first demosaicing, devignetting and MIs alignment [4]. Then, the obtained LLF image (formed by demosaiced, devignetted and aligned MIs, see Figure 1(a)) is rendered into a 15×15 matrix of 2D images (see Figure 1(b)), the so-called *PIs*. In this case, a PI is simply obtained by extracting the pixel with the same position within each MI and putting them all together; the result of this process is a 2D image with a spatial resolution of 625×434 pixels. While 225 (15×15) PIs are originally rendered, both the first and last rows and columns of the PIs 2D array are discarded from further processing (thus resulting into 13×13 PIs) to avoid using the dark PIs associated to the vignetting effect [4][10]. Finally, each PI undergoes color and gamma correction. Note that, due to length constraints, the rendering algorithms for focused cameras are not reviewed here as they are not instrumental to this paper; however, the reader may refer to [11] for a detailed overview.

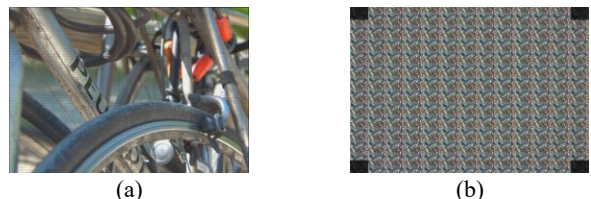

(a)  (b)

Figure 1 – (a) 625×434 matrix of MIs (each with 15×15 pixels); and (b) 15×15 matrix of PIs (each with 625×434 pixels) for the *Bikes* LLF image.

### III. LLFIC: Proposing a Classification Taxonomy

Since multiple technical approaches have been adopted for the LLFIC solutions available in the literature, it is useful to identify their main commonalities, differences and relationships, thus providing a better understanding of the full LLFIC landscape and promising future research and standardization directions. In this context, this paper proposes first a classification taxonomy for LLFIC solutions and will after exercise it by reviewing some of the most relevant LLFIC solutions available in the literature associated to different classification paths. In the next sub-sections, the proposed classification dimensions for the taxonomy will be proposed first. After, the classes for each taxonomy classification dimension will be proposed. The classification dimensions and the classes within each dimension have been defined based on the exhaustive reviewing of tens of LLFIC solutions available in the literature in order a robust taxonomy could be defined [12]-[99]; this list of references is also an useful contribution of this paper.

#### A. Taxonomy Classification Dimensions

This section presents and defines the classification dimensions for the taxonomy proposed for LLFIC solutions. After an exhaustive study of the LLFIC solutions available in the literature, it was concluded that the most appropriate taxonomy classification dimensions are:

1. **Fidelity:** Refers to the fidelity with which the data is coded.
2. **Data Representation Basis:** Refers to the elementary component, i.e. *basis*, in which the raw sensor image data (sensed light intensity and direction information) are represented for coding purposes; depending on the adopted data representation basis, demosaicing, devignetting, alignment, and perspective image (or view) rendering may be involved.
3. **Data Type:** Refers to the type of data that is coded; depending on the adopted data type, depth or disparity estimation may be involved.
4. **Data Structure:** Refers to the way the LLF data, represented in a specific data representation basis, are arranged to be then

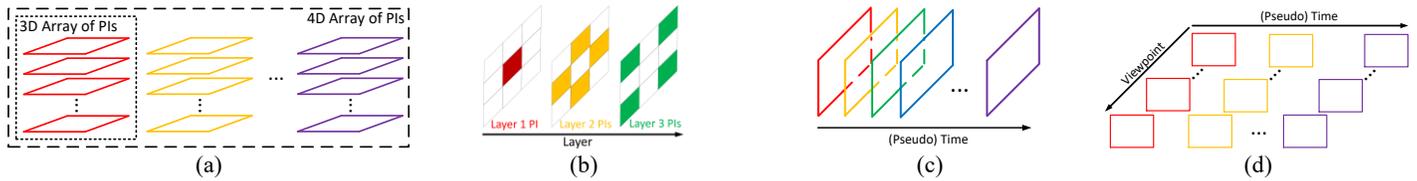

Figure 2 – Illustration of data structure classes: (a) 3D and 4D arrays of PIs; (b) 3-Layered sets of PIs; (c) Pseudo video of PIs; and (d) Pseudo multi-view video of PIs.

coded while exploiting the available spatial and/or angular redundancies.

Using these dimensions, each LLFIC solution may be characterized by a taxonomy *classification path* connecting a set of classes along these dimensions, thus allowing to identify commonalities through the overlapping of the corresponding classification paths. Although the taxonomy does not directly address the coding tools, all the dimensions are directly related to the coding process with a direct impact on the functionalities offered in the relevant application scenarios. For example, the fidelity dimension is directly related to the faithfulness of the data, e.g. critical for medical applications. The data representation basis dimension has a direct relation with rendering, e.g. critical for 2D backward compatibility, stereoscopic and autostereoscopic displaying, interactive navigation and refocusing. The data type and data structure dimensions are directly related to compression efficiency, random access and scalability, e.g. critical for broadcasting, streaming and storage applications to meet the characteristics of multiple types of displays, transmission channels and user needs.

*B. Classes for each Classification Dimension*

Using the proposed classification dimensions, it is now necessary to propose the classes for each classification dimension, naturally based on the LLFIC solutions available in the literature. After exhaustive analysis, the following classes are proposed for each dimension:

1. **Data Fidelity:** In terms of fidelity, the following classes are proposed:
    a. ***Lossless*** – Codecs keeping the original data fidelity, meaning that the decoded and original data are mathematically equal (up to a certain precision, if required).
    b. ***Lossy*** – Codecs not keeping the original data fidelity, typically to increase the compression factor; high fidelity, notably perceptually lossless quality, may still be achieved with the appropriate coding parameters configuration.
2. **Data Representation Basis:** In terms of data representation basis, the following classes are proposed:
    a. ***Micro-Image*** – Representation basis of the LLF data corresponding to the image captured through an individual ML, the so-called *micro-image* (MI). The full LLF image is represented as a set of MIs. Depending on the optical setup used in the LLF acquisition, each MI may capture only angular information (unfocused camera) or both angular and spatial information (focused camera).
    b. ***Perspective Image*** – Representation basis of the LLF data corresponding to an image associated to a specific angular viewing direction of the same scene, the so-called *perspective image* (PI). Depending on the optical setup used in the LLF acquisition, each PI may be obtained by extracting a single pixel or a patch (of pixels) in the same location from each micro-image and putting them together in a 2D array, i.e. a 2D image.
3. **Data Type:** In terms of data type, the following classes are proposed:
    a. ***Texture*** – Information to be coded includes only texture data, i.e. color information associated to the raw sensor data in some color space, e.g. RGB or YUV, and data representation basis.
    b. ***Texture + Geometry*** – Information to be coded includes both texture and geometry-related data, with the latter associated to the 3D arrangement of the scene. The geometry-related data can be either information expressing the distance, measured perpendicularly to the camera's plane, between the camera lens' optical center and the plane containing each scene (3D) point, the so-called *depth*, or information expressing the distance between two 2D image points (pixels) corresponding to the same scene (3D) point projection onto two camera planes, the so-called *disparity*; this information may be used to exploit the available angular redundancy. The geometry-related data may be estimated from the perspective (texture) images.
4. **Data Structure:** In terms of data structure, the following classes are proposed:
    a. ***Single 2D Image*** – The LLF data, represented in a specific basis, are arranged in a 2D array, i.e. an image (see Figure 1). This structure is suitable to be coded with standard-based image and video (Intra mode) coding solutions, e.g. JPEG 2000 or HEVC Intra. In this context, only the spatial correlation within the LLF image (this means the 2D array of MIs) or within the image of PIs (this means the 2D array of PIs) is exploited.
    b. ***Multi-Dimensional Array of Images*** – The LLF data, represented in a specific basis, are arranged in an N-dimensional array of images, usually without specific scanning considerations. The number of array dimensions can be three, this means a stack of either MIs or PIs (see Figure 2(a)), or four, this means a sequence of stacks of PIs (see Figure 2(a)). This structure is suitable to be coded with high-dimensional transform-based coding solutions. In this context, both the LLF spatial (this means within an image) correlation and the inter-view correlation (this means across the images, in one or two directions, depending on the number of array dimensions) are exploited.
    c. ***Layered Sets of Images*** – The LLF data, represented in a specific basis, are arranged in two or more layered sets of images (corresponding to PIs or MIs), as illustrated in Figure 2(b). Altogether, the layered sets of images may correspond to the whole LLF data or only part of it, in case some data are not coded; in the latter scenario, the not coded LLF data

may be obtained at the decoder though image synthesis. Image synthesis, with or without using information from the original LLF image itself, may also be performed at the encoder to create better predictions. This structure is suitable to be coded with a hierarchical coding strategy, allowing different types of coding solutions (standard-based or not) to be applied to the images in each layer (this means hierarchical level). In this context, both spatial correlation (within an image) and inter-view correlation (in one or two directions) are exploited.

d. *Pseudo Video* – The LLF data, represented in a specific basis, are arranged as a 'temporal' sequence of images (corresponding to PIs or MIs) following a specific scanning order (see Figure 2(c)), which targets an increased correlation between adjacent PIs, thus mimicking the high temporal correlation existing in a regular video sequence. This structure is suitable to be coded with standard video coding solutions, e.g. HEVC. In this context, both spatial correlation (within an image) and the correlation between 'temporally' adjacent views (the 'video frames') are exploited.

e. *Pseudo Multi-View Video* – The LLF data, represented in a specific basis, are arranged as multiple 'temporal' sequences of images (typically corresponding to PIs), as illustrated in Figure 2(d). This structure is suitable to be coded with multi-view coding standards, e.g. MV-HEVC or 3D-HEVC. In this context, both spatial correlation (within an image) and the correlation between views (both 'temporally' and angularly adjacent views) are exploited.

An overview of the proposed classification taxonomy is shown in Figure 3; note that the arrows simply intend to highlight example connection paths between classes along the four dimensions. Because the data structure dimension is directly related to compression efficiency, the next section will review in more detail a few, key LLFIC solutions in the literature, to better understand the involved key concepts and designs, guided by the proposed taxonomy's data structure dimension.

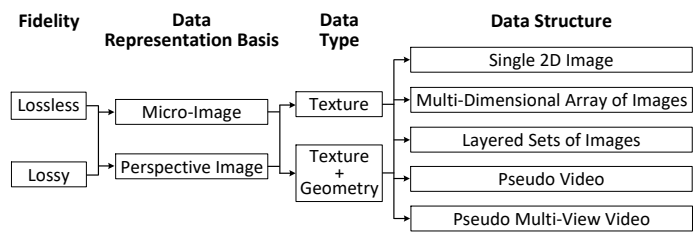

Figure 3 – Overview of the proposed LLFIC classification taxonomy.

## IV. LLFIC: REVIEWING GUIDED BY THE TAXONOMY'S DATA STRUCTURE DIMENSION

To experience and appreciate the power of the proposed classification taxonomy, this section reviews and classifies some relevant and taxonomically representative LLFIC solutions available in the literature guided by the proposed taxonomy's data structure dimension; because these solutions are strategically selected, based on their taxonomical representativeness, relevance and diversity in the LLFIC landscape and performance, a deeper and detailed view of the current LLFIC *status quo* can be obtained. Although this paper's target is not to perform an extensive survey of the LLFIC literature, the authors provide in [100] a summary table where a very large set of LLFIC references are classified according to the proposed taxonomy; this table allows identifying related LLFIC solutions with respect to one or more taxonomy dimensions.

Since taking the human visual system's characteristics into account is a must for efficient image/video coding, most LLFIC solutions in the literature, including those reviewed in this section, perform some perception driven pre-processing before encoding. This pre-processing often involves a conversion from RGB to YUV (4:4:4) color space using some recommendation, e.g. ITU-R BT.709-6, color sub-sampling, e.g. from 4:4:4 to 4:2:0, and bit depth downsampling, from 10-bit to 8-bit.

### A. Single 2D Image based LLFIC

In [76], a LLFIC solution is proposed where the intrinsic correlation between neighboring MIs is exploited by a bi-prediction estimation and compensation tool, so-called *bi-prediction self-similarity* (BI-SS); this tool creates a LLF-biased additional prediction type in HEVC Intra coding, see encoding architecture in Figure 4. The key idea is to complement the powerful HEVC Intra coding tools set with a novel tool designed considering the specific LLF data characteristics when the data representation basis is the MI, naturally targeting a better correlation exploitation. The same HEVC rate-distortion optimization (RDO) process is applied to select the best prediction mode among the HEVC Intra modes and the novel Bi-SS mode, followed by the usual HEVC Intra encoding steps (see Figure 4). Before encoding, the raw LLF image is demosaiced and devignetted (here MIs are not aligned) followed by the pre-processing steps described at the beginning of Section IV. The resulting 8-bit YUV 4:2:0 2D (LLF) image constitutes the input for the encoder, hereafter called *Bi-SS encoder*.

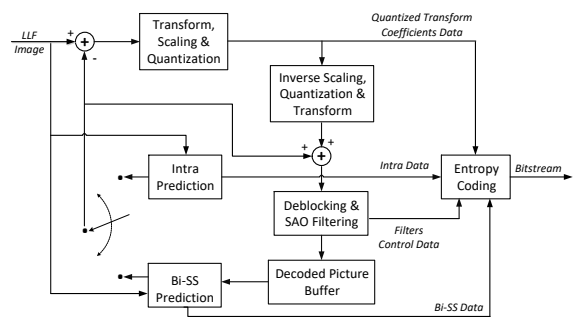

Figure 4 – Bi-SS LLF image encoding architecture.

The main Bi-SS codec encoding steps involve:

1. **HEVC Intra Prediction:** First, the 35 HEVC Intra prediction modes are evaluated for all possible Intra block sizes and the mode leading to the lowest RD cost (according to the HEVC RDO process) is selected as the best HEVC Intra prediction mode.

2. **Bi-SS Prediction:** For each coding block (CB) size, ranging from 64×64 to 8×8, two SS-based prediction candidates are obtained from the same LLF image, through full search within the same causal neighborhood (formed by decoded blocks) of the CB to be predicted: 1) the uni-prediction SS (Uni-SS) candidate, corresponding to a prediction block estimated as in HEVC uni-predictive (P frame) coding but using as reference the LLF image itself (instead of a preceding image), the so-called *SS reference*; and 2) the bi-prediction SS (Bi-SS) candidate, which corresponds to a weighted combination of two prediction blocks, jointly estimated from the same full (causal) search area of the (same) SS reference. The joint estimation of the two prediction blocks is performed through an iterative process by minimizing a Lagrangian cost function conditioned to the optimal prediction found in the previous iteration; the Uni-SS prediction candidate is used as the starting point for this iterative process. After the joint estimation iterative process reaches a certain number of iterations, the HEVC RDO process is applied to select the best prediction between the Uni-SS and Bi-SS candidates. To select the final prediction, the HEVC RDO process is applied again between the (best) SS prediction and HEVC Intra prediction.

3. **Transform, Scaling & Quantization:** The prediction residue between the CB and the best prediction previously found is then computed. This prediction residue is after transformed, scaled and quantized as in regular HEVC Intra coding.
4. **Inverse Scaling, Quantization & Transform:** The quantized transform coefficients are then inversely scaled, quantized and transformed, thus obtaining the decoded prediction residue, which is then added to the appropriate prediction to reconstruct the image samples.
5. **Deblocking & SAO Filtering:** The deblocking and sample adaptive offset (SAO) filters are applied over the decoded image samples to reduce the coding-related artifacts. The resulting filtered LLF image is stored in the decoded picture buffer to be used as reference for Intra/Bi-SS prediction.
6. **Entropy Coding:** Finally, the coding bitstream correspondent to the coded LLF image is obtained by applying context adaptive binary arithmetic coding to the quantized transform coefficients data stream and the syntax elements, naturally also including the new SS prediction modes signaling.

In terms of the proposed classification taxonomy, the presented Bi-SS LLF codec corresponds to a *Lossy–Micro-Image–Texture–Single 2D Image* path. As mentioned above, the Bi-SS coding solution [76] considers as input a LLF image represented as a 2D array of MIs whose centers have not been aligned. This implies the decoder will need to receive some metadata, such as the MI centers, the MIs size and color calibration data, in order it is possible to render the corresponding 2D array of PIs, thus obtaining the output data for display, e.g. in the format defined in the JPEG Pleno LFC CTC [10]. However, no metadata coding solution has been considered in [76], thus implying that the total rate does not account the metadata required to produce 2D rendered images if alignment has to be performed at the decoder. Thus, to allow a fair and meaningful performance comparison with other LLFIC solutions when final 2D displaying is targeted, the Bi-SS coding solution evaluated in Section V.D will consider as input a LLF image formed by already aligned MIs (see Figure 1(a)), thus avoiding the need for metadata coding.

### B. Two-Layered Sets of Images based LLFIC

In [87], a LLFIC solution is proposed where the relationship (or correlation) between the full set of original PIs (rendered from the raw LLF image as described in Section II) is exploited through a graph-based representation. A graph is a data structure characterized by a set of nodes/vertices and corresponding connections/edges. The key idea is to estimate how similar each PI (corresponding to a graph node/vertex) is to each of the remaining PIs (in the full set of original PIs) and represent that similarity as a connection weight between two PIs. This allows defining a framework where, by combining the PIs according to the respective weights, it is possible to interpolate any PI within the full set of PIs. To take advantage of this graph-based representation to achieve high compression efficiency, this graph-based (GB) LLFIC solution [87] organizes the full set of (original) PIs in two layers, as depicted in Figure 5, thus allowing to selectively process each layer in a different way.

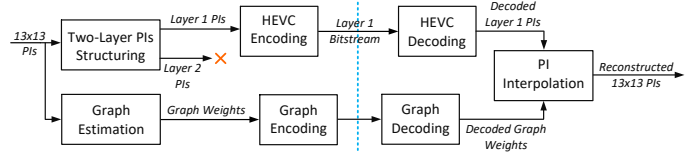

Figure 5 – GB LLF image coding architecture.

The main GB codec coding steps involve:

1. **Two-Layer PIs Structuring:** After applying the pre-processing steps described at the beginning of Section IV (except the bit depth downsampling), the resulting 10-bit YUV 4:2:0 (13×13 central) PIs are arranged in a two-layered structure, following a chess pattern-like data split. In this context, the first layer (L1) will contain 85 PIs out of 169, while the second layer (L2) will contain the remaining 84 PIs, which will be interpolated at the decoder based on the decoded L1 PIs and the graph weighs.
2. **HEVC Encoding/Decoding:** The 10-bit YUV 4:2:0 L1 PIs are organized into a pseudo video sequence, following a serpentine scanning order, and encoded with the HEVC Main10 profile and low delay configuration (IPP...) [101].
3. **Graph Estimation:** A graph is estimated based on the 13×13 central, original PIs. Basically, the luminance component of each PI is mapped to a node in a weighed graph, i.e. a graph with weights associated to its edges. Then, the graph weighted adjacency matrix $W$, with weights associated to the graph edges, representing the similarity between the two vertices (in this case, two PIs) it connects, is obtained through a graph learning technique based on weighted L-1 norm minimization [87].
4. **Graph Encoding/Decoding:** The weights associated to the graph edges ($W$) are then coded without compression using 8 bytes to represent each weight. The elementary encoded bitstreams resulting from the first layer, notably HEVC coded PIs, and the graph weights, are multiplexed, thus generating the final LLF coding bitstream.
5. **PI Interpolation:** Since the graph represents the relationship between the full set of (original) PIs, by knowing the graph weights along with the L1 decoded PIs, L2 PIs are obtained as a weighted combination of the L1 decoded PIs [87]:

$$\hat{X} = (M + \gamma L)^{-1}\hat{Y}. \qquad (1)$$

In (1), $\hat{Y}$ is a matrix where each row corresponds to a 1D-vectorized decoded PI (or zeros in case the corresponding PI is L2), $L$ is the graph Laplacian matrix obtained from the decoded graph weights, and $M$ is an identity matrix with zeros on the diagonal indices corresponding to L2 PIs. While L2 PI interpolation is performed based on decoded L1 PIs (and graph weights), the graph estimation is performed based on the full set of original PIs. This original-decoded mismatch may lead to a loss in the reconstructed L2 PIs quality, which is expected to be more evident for lower bitrates, where the L1 PIs quality is more significantly affected by the compression operations. This effect is reduced by introducing the parameter $\gamma$ in (1), where it acts as a compensation mechanism for the L1 PIs quality mismatch (between original at encoder and decoded at decoder). The reconstruction process in (1) is performed independently for the Y, U and V components, with a different $\gamma$ value (empirically obtained) for the luminance and chrominances, although always using the same graph weights, which have been estimated from the full set of original PIs luminance.

In terms of the proposed classification taxonomy, the presented GB LLF codec corresponds to a *Lossy–Perspective Image–Texture+Geometry–Layered Sets of Images* path. Differently from the coding solution described in Section IV.A, which exploits the correlation between MIs, the GB LLF codec exploits the correlation within (Intra) and between (Inter) PIs, which may allow a more flexible exploitation of the redundancy as it is not constrained by the MLA structure.

### C. 4-Dimensional Array of Images based LLFIC

In [89], a so-called *Multidimensional Light field Encoder using 4D Transforms and Hexadeca-trees* (*MuLE-TH*) LLF image coding solution is proposed, where the intrinsic spatial-angular or 4D LLF redundancy is exploited as a whole by means of a 4D transform, see architecture in Figure 6. The key idea is to take advantage of a tree data structure based transform coefficients' bitplanes decomposition to attain a more efficient compression.

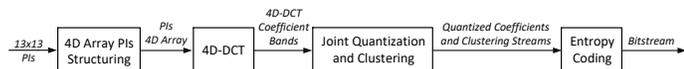

Figure 6 – MuLE-TH LLF image encoding architecture.

The MuLE-TH encoding process proceeds as follows:

1. **4D Array PIs Structuring:** After applying the pre-processing steps described at the beginning of Section IV (except the bit depth downsampling), the resulting 10-bit YUV 4:2:2 (13×13 central) PIs are arranged in a 4D array structure; the RGB to YUV color space conversion has been performed using recommendation ITU-R BT.601-5.

2. **4D-DCT:** Both the luminance and chrominance components of the 4D array of PIs are divided, one component at a time, into $N_A \times N_A \times N_S \times N_S$ (i.e. 4D) blocks, where $N_A$ and $N_S$ correspond to the block sizes in the angular (or inter-PI) and spatial (intra-PI) LLF dimensions, respectively. Then, a 4D separable Discrete Cosine Transform (DCT) is applied over the 4D blocks samples; this means that an 1D-DCT is applied to each block dimension, one at a time, starting by the two angular or inter-PI dimensions, followed by the two spatial or intra-PI dimensions. The 4D-DCT transform coefficients are then grouped into a 4D array of sub-bands, according to the position occupied by each DCT coefficient within the 4D blocks.

3. **Joint Quantization and Clustering:** The 4D-DCT coefficients are clustered or grouped using a so-called *hexadeca-tree* data structure where each node corresponds to a 4D block of DCT coefficients for a given sub-band. Each node can be further divided into sixteen children nodes or not depending on the respective DCT coefficients *significance*, which is determined on a bitplane basis, starting with the most significant one; bitplanes are generated by converting the 4D-DCT coefficients (represented by integer values) into a binary representation. A 4D-DCT coefficient is considered *non-significant* at bitplane $B$ if its bit in bitplane $B$ and the bits corresponding to the bitplanes more significant than $B$ are all zero; otherwise, the 4D-DCT coefficient is considered *significant*. When a 4D-DCT coefficients block (for a given sub-band) has only *non-significant* (i.e. zero) 4D-DCT coefficients or a minimum size of 1×1×1×1, it will not be divided and a '0' is outputted to signal that occurrence; otherwise, the 4D-DCT coefficients block will be further divided into sixteen sub-blocks and a '1' is output. Whenever the block reaches the 1×1×1×1 size, the corresponding DCT coefficient value, represented by its $(30 - L)$ most significant bits, is sent to the entropy encoder; while $2^{30}$ is the pre-set value for the maximum allowed DCT coefficient amplitude, $2^L$ corresponds to the quantization step size.

   The hexadeca-tree decomposition of the 4D-DCT coefficients bitplanes successively proceeds from the most significant bitplane to the least significant one, where the latter is defined by the target quantization step size; a quantization step size of $2^L$ implies that the hexadeca-tree decomposition is performed until $B = L$. By traversing the hexadeca-tree, a bitstream is obtained at the bitplane level, where '1' ('0') indicates that the respective hexadeca-tree node has been divided (has not been divided). The hexadeca-tree decomposition of the 4D-DCT coefficients bitplanes allows, therefore, efficiently representing 4D blocks with only non-significant (or zero) DCT coefficients with a single 0 symbol for each block, while localizing the significant ones.

   Besides the bitstream resulting from the hexadeca-tree traversing (indicating the locations of the non-zero coefficients), this module also outputs the (quantized) DCT coefficients which survived the hexadeca-tree decomposition, this means the 4D-DCT coefficients residing in a 1×1×1×1 block with amplitude larger than 0.

4. **Entropy Coding:** Finally, this module creates a bitstream exploiting the statistics of all the data output by the previous module, i.e. the hexadeca-tree partitions and (quantized) DC and AC coefficients streams. This entropy encoder takes into account the specific symbol frequencies for the various data streams to be entropy coded. In this solution, a context-based binary adaptive arithmetic coder is used with three different symbol frequency tables, one for each data streams aforementioned, i.e. hexadeca-tree partitions and (quantized) DC and AC coefficients streams. This module outputs the coding bitstream corresponding to the coded LLF.

In terms of the proposed classification taxonomy, the presented MuLE-TH LLF codec corresponds to a *Lossy–Perspective Image–Texture–Multi-Dimensional Array of Images* path. Differently from the coding solution described in Section IV.B, which exploits the inter-PI and intra-PI correlations separately and using different tools, the MuLE-TH codec exploits the intrinsic 4D LLF correlation as a whole using the 4D-DCT, which may allow a more efficient exploitation of the redundancy. The MuLE-TH codec has meanwhile been improved [5] regarding in initial description [89], both in terms of compression efficiency and random access capability, notably by adopting RD-optimized 4D block partitioning and quantization strategies, and independent coding of each 4D block; 10-bit YUV 4:4:4 PIs are supported at the encoder input. The improved MuLE-TH codec, hereafter called *MuLE*, has been adopted as the *4D transform mode* in the JPEG Pleno standard, due to its good performance in terms of compression efficiency and random access for LLF content [99]. For this reason, the coding solution evaluated in Section V.D will correspond to the JPEG Pleno standardized version of the MuLE codec.

### D. N-Layered Sets of Images based LLFIC

In [94], a so-called *Warping and Sparse Prediction* (*WaSP*) based LLFIC solution is proposed, which is based on a layered arrangement of the full set of PIs (rendered as described in Section II), see encoding architecture in Figure 7. The key idea is to predictively encoded each PI layer from (previously encoded) PIs belonging to lower layers, thus exploiting the correlation between neighboring (or nearby) PIs while providing random access. Since the first (and lowest) layer of PIs does not have previous layers, the first layer PIs are encoded independently, i.e. without exploiting any inter-PI correlation; the WaSP solution also encodes depth data for the first layer PIs, which are used to provide the higher layers the appropriate information needed to compensate the disparity between PIs, i.e. to generate warped PIs which allow creating predictions for higher layers PIs.

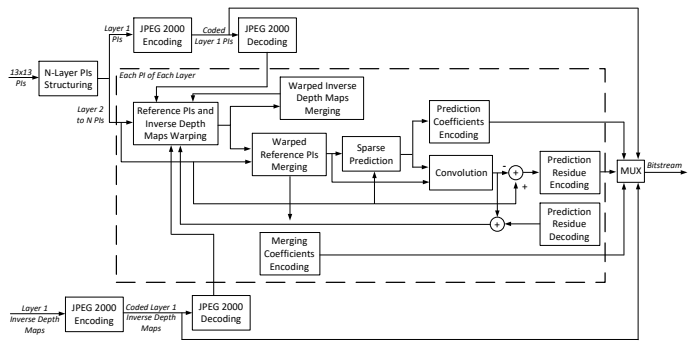

Figure 7 – WaSP LLF image encoding architecture.

The WaSP encoding process proceeds as follows:

1. **N-Layer PIs Structuring:** The 10-bit RGB 4:4:4 (13×13 central) PIs are first arranged into a N-layered structure. While the WaSP solution defines six as the maximum number of (encoding) layers, this value can be adjusted according to the application requirements. These layers (except the last one, L6) are also known as *reference layers* because the corresponding PIs are used as reference PIs for the prediction of higher layers PIs.

*Layer 1*

2. **JPEG 2000 Encoding/Decoding:** All the L1 data is coded with the same coding solution, this means JPEG 2000:
   - **L1 PIs** – First, the 10-bit RGB L1 PI (the central PI in this case) is coded with JPEG 2000 [102][103] to be used as reference for the PIs in higher layers.
   - **L1 Inverse Depth Maps** – An inverse depth map, whose values correspond to the ratio between the camera focal length and the depth value at each pixel location in L1 PI, is also coded with JPEG 2000; this map will be used later to synthesize inverse depth maps for PIs in higher layers. The inverse depth map is estimated based on the occlusion-aware depth estimation method proposed in [104].

*Layers 2 to 6*

3. **Reference PIs and Inverse Depth Maps Warping:** Given the inverse depth map for the reference PIs, the inverse depth map for any other PI in higher layers, so-called *warped inverse depth map*, is synthesized by a simple pixel-wise warping operation of the inverse depth map values associated to the reference PIs [94]. The corresponding reference PIs are then warped to the location (or viewing angle) of the PI to code by copying the intensity value at position (*i*, *j*) of the reference PI to the corresponding warped position of the so-called *warped reference PI*.

4. **Warped Reference PIs Merging:** The warped reference PIs are fused to generate a single high-quality image, so-called *merged PI*, to be used in the prediction stage (Step 7). In this case, each merged PI pixel results from the fusion of pixel values from different warped reference PIs, depending on whether the (PI to code) pixel is or not visible on the reference PI. The optimal contribution of each warped reference PI to the merged PI is determined through the popular least-squares (LS) regression method [94].

5. **Warped Reference Inverse Depth Maps Merging:** The warped inverse depth maps are also merged. In this case, a simple median strategy is adopted whenever multiple warped inverse depth values are obtained for the same merged inverse depth map pixel position; when a single warped inverse depth value is obtained for a given merged inverse depth map pixel position, this is the value used for that position.

6. **Merging Coefficients Encoding/Decoding:** For the decoder to perform the warped reference PIs merging operation, the LS regression model parameters ($\theta$) determined in Step 4 have to be coded. In this case, the $\theta$ parameters are coded without compression using 2 bytes.

7. **Sparse Prediction:** To predict the PI to code from the merged PI obtained in Step 4, a sparse linear prediction model is built for each color component, which are sequentially encoded. The model $\Theta$ prediction coefficients are estimated using the LS method to minimize the error between the color component to be predicted and the estimate given by the model.

8. **Prediction Coefficients Encoding/Decoding:** This module codes vector $\Theta$, which corresponds to the non-zero prediction coefficients values, by applying arithmetic coding to the binary string representing the non-zero coefficients location and Golomb-Rice coding to the amplitude of the predictions coefficients.

9. **Convolution:** After obtaining the predictions, the prediction residue is simply obtained by subtracting a certain color component of each pixel in the PI to code from the prediction, which in this case is obtained by convolving the merged PI with a specific sparse predictor, notably the one obtained in Step 7 after the quantization operation above mentioned.

10. **Prediction Residue Encoding/Decoding:** The prediction residue is coded with JPEG 2000; however, a more efficient coding solution can be used. The elementary encoded bitstreams resulting from the JPEG 2000 encoded prediction residue, the encoded prediction coefficients, the encoded merging reference PIs coefficients, and the JPEG 2000 encoded first layer (both texture and inverse depth maps) are multiplexed to generate the final coding bitstream.

In terms of the proposed classification taxonomy, the presented WaSP based LLF codec corresponds to a *Lossy–Perspective Image–Texture+Geometry–Layered Sets of Images* path. As the MuLE-TH (and MuLE) LLF image codec, the WaSP based LLF codec also exploits 4D LLF correlation, in this case through the (4D) sparse prediction. The WaSP solution has been adopted as the *4D predictive mode* in the JPEG Pleno standard and has been recognized as more efficient than the MuLE solution (aka 4D transform mode), also included in JPEG Pleno, for high density camera arrays LFs [5].

### E. Pseudo Multi-View Video based LLFIC

In [90], a LLFIC solution is proposed where the 2D array of PIs (see Section II) is represented as a pseudo multi-view (PMV) video sequence, see encoding architecture in Figure 8; this solution will be hereafter referred as *PMV solution*. The key idea is to take advantage of conventional multi-view coding standards, in this case MV-HEVC, to exploit the intrinsic 4D LLF correlation. For that purpose, a 2D prediction structure is designed to better adapt to the PIs 2D array structure, thus enabling a more efficient spatial-angular redundancy exploitation. As in MV-HEVC, the 2D prediction structure is used to define the coding dependencies between frames (or PIs in this case) of multiple views as well as between video frames in the same view; this is done by assigning to each view frame (i.e. PI) a so-called *prediction level*. In other words, PIs assigned with a given prediction level are predictively encoded using as references previously decoded PIs with lower prediction levels, thus exploiting the correlation between PIs. In the PMV solution, the prediction structure is also used to determine the quantization parameter (QP) to be applied to some PIs. In this process, the MV-HEVC coding process itself is not changed at all.

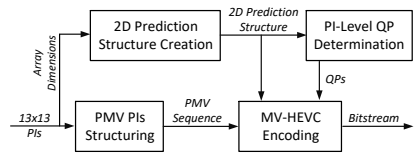

Figure 8 – PMV LLF image encoding architecture.

The main PMV codec encoding steps involve:

1. **PMV PIs Structuring:** After applying the pre-processing steps described at the beginning of Section IV, the resulting 8-bit YUV 4:2:0 (13×13 central) PIs are first arranged in a PMV video sequence. The PMV video sequence includes 13 views, corresponding to the PIs 2D array 13 rows, with 13 frames each, corresponding to the PIs 2D array 13 columns.
2. **2D Prediction Structure Creation:** The creation of the 2D prediction structure is done through an iterative process, for both the horizontal and vertical directions. Thus, for each row/column of the PIs 2D array, starting with the middle one, the central PI is assigned to the lowest prediction level ($P_0$); the first and last PIs within that row/column are also assigned to $P_0$. Then, the PI in the middle of two consecutive PIs with a prediction level already assigned is assigned to the next prediction level. This process is repeated while the number of PIs in-between two consecutive PIs with a prediction level already assigned is higher than or equal to 2; the remaining PIs will be assigned to the highest prediction level, meaning that they will not be used as reference frames for prediction. Finally, the first and last PIs within the row/column are assigned to prediction level 1.
3. **PI-Level QP Determination:** The QP for each PI is obtained by adding an offset to a reference QP, which is assigned to the central PI. For the PIs located in the central row/column of the PIs 2D array (except the central PI), the QP offset is set to the maximum prediction level value of that PI in the horizontal and vertical directions. For the remaining PIs, the QP offset is computed based on the PI distance and view-wise decoding order with respect to the PI with the lowest prediction level.
4. **MV-HEVC Encoding/Decoding:** The 8-bit YUV 4:2:0 PMV sequence is coded with MV-HEVC using the 2D prediction structure and the QP values previously obtained. The resulting encoded bitstream corresponds to the coded LLF image.

In terms of the proposed classification taxonomy, the presented PMV based LLF codec corresponds to a *Lossy–Perspective Image–Texture–Pseudo Multi-View Video* path. Differently from the WaSP codec (see Section IV.D), which exploits the 4D LLF correlation through 4D sparse prediction, the PMV solution exploits the 2D angular (inter-PI) and 2D spatial (intra-PI) correlation separately and selects the best one in terms of RDO. This may allow a more efficient exploitation of the intrinsic spatial-angular redundancy, since the prediction block sizes may be independently adjusted to the amount of 2D spatial and 2D angular correlations available for exploitation.

## V. LLFIC: PERFORMANCE ASSESSMENT

In the previous section, some relevant LLFIC solutions available in the literature have been reviewed at the light of the proposed taxonomy to exercise and demonstrate its potential. To have a deeper understanding of the LLFIC field, a direct comparative performance analysis is provided in this section. For this comparison to be fair, it is essential to select meaningful and precise test conditions and evaluation metrics. Considering the context, the natural choice for these test conditions and evaluation metrics are the JPEG Pleno Light Field Coding (LFC) Common Test Conditions (CTC) [10] as they have been defined by the most relevant standardization group in the LLFIC arena.

### A. Test Material and Conditions

The JPEG Pleno LFC CTC test material and conditions for LLFIC have been adopted in this paper and are briefly summarized here [10]:

- **Test material:** *Bikes, Danger de Mort, Stone Pillars Outside* and *Fountain&Vincent2* (the central PI for each LLF image is shown in Figure 9); these LLF images have been acquired with the Lytro Illum (10-bit) LF camera (an *unfocused camera*) and represent natural and outdoor content with objects at different depths. These LLF images are part of the JPEG Pleno database, publicly available at [105].
- **Input and output components:** Test LLF images are in portable pixmap (PPM) file format, i.e. 10-bit images with non-interlaced RGB color components, and the output LLF images must also be provided in 10-bit PPM file format. For performance evaluation purposes, both input and output components (10-bit PPM) are converted to (10-bit) YUV 4:4:4 color space using the ITU-R BT.709-6 recommendation. However, within the codec, the LLF images may be coded using any color space and bit depth.
- **Number of PIs:** Only the central 13×13 PIs, out of the total 15×15 PIs, are used for coding and performance evaluation purposes to avoid using the rather dark PIs associated to the vignetting effect, corresponding to the top and bottom rows and the most left and most right columns.

- **Spatial resolution:** Each PI contains 625×434 pixels: This resolution may need to be adjusted to obtain a spatial resolution more compatible with some standard coding solutions, e.g. multiple of 8 or 16; however, this adjustment only applies to the coding solutions in the *Perspective Image* class.
- **Target bitrates:** 0.001, 0.005, 0.02, 0.2 and 0.75 bits per pixel (bpp); these values allow covering a wide range of bitrates (and qualities) for different content characteristics. The two lowest bitrates may be unachievable for some standard coding solutions, as they would require using a QP value higher than the maximum allowed.

These conditions play a key role in a meaningful direct comparison of LLFIC solutions since most performance results in the literature are not effectively comparable as following different conditions.

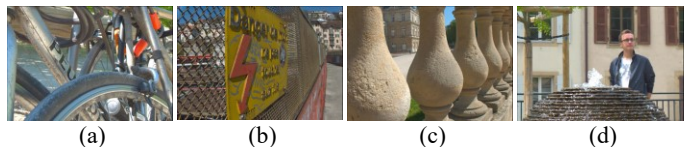

Figure 9 – Central PI for each LLF image: (a) *Bikes*; (b) *Danger de Mort*; (c) *Stone Pillars Outside*; and (d) *Fountain&Vincent2*.

### B. Evaluation Metrics

The JPEG Pleno LFC CTC evaluation metrics for LLFIC have also been adopted in this paper and are briefly summarized here [10]:

- **Bitrate:** Number of bits per pixel (bpp), which is defined as the ratio between the number of bits for the coded LLF representation and the number of pixels in the whole LLF; for encoding a LLF composed by 13×13 views (or PIs), each with 625×434 pixels, the number of pixels in the whole LLF is 13×13×625×434 = 45 841 250.
- **Objective quality:** Average peak signal-to-noise ratio (PSNR) and average structural similarity index (SSIM) [106] of the luminance component for the whole LLF, represented as PSNR-Y and SSIM-Y, respectively, and weighted average of the individual PSNR components, i.e. PSNR-Y, PSNR-U and PSNR-V, for the whole LLF, represented as PSNR-YUV. The average PSNR-Y/SSIM-Y/PSNR-YUV for the whole LLF is computed as the average of the PSNR-Ys/SSIM-Ys/PSNR-YUVs for the 13×13 PIs [10].
- **Bjøntegaard-deltas:** Bjøntegaard delta (BD) rate (BD-rate) and BD-PSNR [107], which measures the coding efficiency gains in rate and PSNR of a specific LLFIC solution with respect to a reference one, in this case the HEVC Main10 profile (see Section V.C).

### C. Anchor Coding Solutions

In this section, the defined JPEG Pleno LFC CTC coding anchor, i.e. HEVC Inter [10], and other three relevant anchors are used for benchmarking. Following the approach adopted by JPEG Pleno LFC CTC [10], all coding anchors presented here use a 4:4:4 chrominance format; although this is not the most efficient format in terms of coding, it is used in this context just to establish a reference for comparison. The four anchor coding solutions used for benchmarking are the following:

- **JPEG 2000:** This image codec exploits the spatial correlation through a (2D) discrete wavelet transform. When coding with JPEG 2000, the 13×13 (10-bit PPM) PIs are coded, all at once, as a single 2D image; this allows a better exploitation of the spatial correlation than individually coding the PIs as the 2D wavelet transform is performed over the whole image. The software used for JPEG 2000 coding was OpenJPEG software, version 2.3.0 [108]. To obtain different RD points, the compression ratio parameter was adjusted to reach the target bitrates defined in the JPEG Pleno LFC CTC [10].
- **HEVC Intra/Inter:** This video codec is the best performing standard Intra/Inter video coding solution. It adopts a hybrid coding architecture [109] to exploit both the spatial and temporal correlations. The JPEG Pleno LFC CTC adopted the HEVC Main10 profile as anchor for LLFIC [10]. When coding with HEVC Inter, the 13×13 (10-bit YUV 4:4:4) PIs are coded as a pseudo video sequence, which is generated by scanning the 13×13 PIs following a serpentine scanning order [10]. When coding with HEVC Intra, the 13×13 (10-bit YUV 4:4:4) PIs are coded as a single 2D image, similarly to JPEG 2000; this may allow a better exploitation of the spatial correlation, particularly around the PI borders. The software used for HEVC coding was x.265 software, version 2.3 [110]. To obtain different RD points, the rate was adjusted to match the target bitrates defined in the JPEG Pleno LFC CTC [10].
- **VVC Intra:** This video codec is currently under development by the ITU-T JVET and ISO/IEC MPEG [111] and targets becoming the next generation video coding standard, to be known as Versatile Video Coding (VVC). As in HEVC, also VVC adopts a hybrid coding architecture to exploit both the spatial and temporal correlations, using again more powerful tools. When coding with VVC Intra, the 13×13 (10-bit YUV 4:4:4) PIs are coded as a single 2D image, similarly to HEVC Intra. The software used for VVC coding was the VTM reference software, version 4.0.1 [112]. To obtain different RD points, the quantization parameter was adjusted to reach the target bitrates defined in the JPEG Pleno LFC CTC [10]. Performance results are not reported here for VVC Inter coding because, at the time this paper was written, the VTM reference software did not support YUV 4:4:4 Inter coding, which is the color sub-sampling format used for all the previously presented anchor solutions.

While patent licensing costs for a codec may be affordable for some application domains, for others, they may act as an inhibitor for codec adoption. Traditionally, standard image codecs are not burdened by royalties while standard video codecs are heavily burdened. Thus, to take the licensing issues into consideration, both royalty-free (JPEG 2000) and non-royalty-free (HEVC and VVC) coding anchors have been considered in this paper with the understanding that, currently, better compression performance is offered by the royalties burdened coding solutions. It is also worth noting that the performance results for all the anchor coding solutions above presented have been obtained by the authors.

### D. Comparative RD Performance Assessment

This section reports the RD performance for the five LLFIC solutions reviewed in Section IV, namely Bi-SS, GB, MuLE, WaSP and PMV, according to the test conditions described above. This comparative RD performance analysis will be driven by the codecs licensing type, i.e. royalty-free versus non-royalty-free[2], due to the current importance of licensing issues on the design and adoption of coding solutions. In practice, the licensing type is also a 'frontier' between JPEG and MPEG, which typically follow the royalty-free and the non-royalty-free models as shown by the set of anchors above. From the five LLFIC solutions reviewed in Section IV, Bi-SS, GB and PMV are classified as non-royalty-free solutions while MuLE and WaSP are classified as royalty-free following their usage or not of non-royalty-free coding solutions, notably based on HEVC which is heavily royalties burdened. In the following, the RD performance results will be first analyzed within each licensing type, royalty free and non-royalty-free, and after the best LLFIC solutions from each type will be compared. Due to space constrains, only PSNR-YUV and SSIM-Y results are reported here; the full set of RD performance results may be found in [100].

*1) Royalty-Free LLFIC Solutions Assessment*

Figure 10 shows the PSNR-YUV RD performance charts for the selected royalty-free LLFIC solutions, namely MuLE, WaSP and the JPEG 2000 anchor. From the charts, the following conclusions may be draw:

- **Royalty-free LLFIC vs. (Royalty-free) JPEG 2000:** The MuLE and WaSP RD performances are consistently and considerably better than the JPEG 2000 RD performance, for all LLF images and bitrates, with average BD-rate savings of 96.3% and 93.8%, respectively; the average BD-rate savings are computed over the four selected LLF test images. This is an expected behavior due to the JPEG 2000 lack of capability in exploiting the LLF intrinsic spatial-angular redundancy, which is significantly different from the usual (spatial) redundancy present in natural images.
- **MuLE vs. WaSP:** The MuLE RD performance is better than the WaSP RD performance, for all LLF images and bitrates, with average BD-rate savings of 29.4%; the highest BD-rate saving is observed for *Bikes*, which is one of the LLF images with the smaller inter-view correlation [113][10]. In general, PSNR gains (measured between two RD points with similar bitrate belonging to two RD curves) decrease when the bitrate increases. This behavior may be explained by the fact that the MuLE codec exploits the 4D redundancy mainly through a 4D block-based DCT while the WaSP codec relies on a hierarchical coding scheme with (sparse) view prediction. As the bitrate increases, the view prediction quality for the WaSP codec increases, since PIs (angularly) closer to the one to be coded are used for the prediction estimation [94]. This translates into a lower prediction residue to be encoded and, consequently, in bitrate savings for similar quality. The MuLE codec follows a simpler strategy by applying always the same adaptive block-size 4D block-based transform, thus exploiting the 4D redundancy in the same way, independently of the bitrate.

*2) Non-Royalty-Free LLFIC Solutions Assessment*

Figure 11 shows the PSNR-YUV RD performance for the selected non-royalty-free LLIFC solutions, namely Bi-SS, GB, PMV and the defined JPEG Pleno LFC CTC anchor, i.e. HEVC Inter [10]; the remaining non-royalty-free anchors (HEVC Intra and VVC Intra) will be considered in the following best LLFIC solutions comparison (see Section V.D.3). From these charts, the following conclusions may be obtained:

- **Non-royalty-free LLFIC vs. (Non-royalty-free) HEVC Inter:** In general, the GB and PMV RD performances are better than the HEVC Inter RD performance, for all LLF images and bitrates, with average BD-rate savings of 40.1% and 49.5%, respectively; the only exception is for the GB codec when coding *Fountain&Vincent2* at the highest bitrate. Moreover, for both GB and PMV codecs, PSNR gains increase when the bitrate decreases, for all LLF images. This behavior indicates that HEVC Inter has more difficulty in obtaining good predictions (thus low residues to code), especially at very low bitrates, since it exploits the inter-PI redundancy only in one angular dimension, while the PMV and GB codecs exploit the inter-PI redundancy in both the angular dimensions. The Bi-SS codec outperforms the HEVC Inter codec for the lowest and highest bitrates, for all LLF images. This trend is not observed for the medium bitrates and varies with the LLF images content, with the highest coding losses observed for *Fountain&Vincent2*, the LLF image with the smallest inter-view correlation [113][10]. This behavior seems to indicate that, as long as the bitrate is not too small, HEVC Inter reaches medium qualities sooner than Bi-SS but its RD performance also saturates quicker that the Bi-SS RD performance; this is possibly due to the HEVC Inter difficulty in obtaining good predictions as the rate increases allied to the less efficient spatial-angular redundancy exploitation, resulting from considering only one angular dimension. Moreover, all the non-royalty-free LLFIC solutions use apply a 4:2:0 color sub-sampling to the 4D LLF before encoding; this may lead to a color fidelity loss when compared to HEVC Inter, where no color sub-sampling has been applied [10].
- **Bi-SS vs. GB vs. PMV:** First, comparing the LLFIC solutions in the PI class of the data representation basis dimension, it can be noticed that the PMV RD performance is consistently better than the GB RD performance, for all LLF images and bitrates, with average BD-rate savings of 15.5%. This behavior indicates that the spatial-angular redundancy is more efficiently exploited through 2D inter-view prediction than using only 1D inter-view prediction with disparity compensation at the decoder (to help reconstruction the full 4D LLF); this is particularly evident at the highest bitrate for the LLF images with the smallest inter-view correlation, i.e. the *Fountain&Vincent2* and *Bikes* LLF images [113][10]. This behavior may be explained by the fact that, the GB solution exploits the inter-PI redundancy in a global way, i.e. at the PI level (each graph vertex represents an entire PI), while the PMV codec does it locally, i.e. at a coding block level, thus becoming more adaptive. Comparing the Bi-SS RD performance, which corresponds to a MI data representation basis, with the GB and PMV RD performances (PI data representation basis), it can be observed that the former is consistently worse, for all LLF images, and most bitrates; the only exception is the highest bitrate, where the Bi-SS RD performance is always better than the GB RD performance.

---

[2] Disclaimer: The "royalty-free"/"non-royalty-free" terminology used in Section V does not have any legal value and is simply based on common knowledge in the coding community, past (MPEG and JPEG) legacy/tradition and commitments made within JPEG. While it is well known that MPEG standards like H.264/AVC and HEVC are heavily burdened by royalties, it is also know that JPEG standards are traditionally royalty-free or have at least a royalty-free baseline codec/profile. Based on this rather consensual knowledge, it is reasonable to adopt the proposed "royalty-free"/"non-royalty-free" classifications for the selected LLFIC solutions as this is a critical element for the full understanding of the LLFIC landscape.

This behavior indicates that the spatial-angular redundancy can be more efficiently exploited when the LLF data is represented in a PI basis than in a MI basis. This may be explained by the difficulty in obtaining good (spatial) predictions when the spatial characteristics of the image to code differ from the usual ones present in natural images; although the Bi-SS codec has a tool specifically designed to exploit the inter-MI redundancy, it still mostly relies on the HEVC Intra prediction tools.

*3) Best Royalty-Free versus Non-Royalty-Free LLFIC Solutions Comparison*

Figure 12 shows the PSNR-YUV RD performance for the best performing royalty-free and non-royalty-free LLFIC solutions, namely MuLE and PMV, and all the LLFIC anchor coding solutions presented in Section V.C. From these charts, the following conclusions may be derived:

- **LLFIC solutions vs. Royalty-free coding anchor:** The MuLE and PMV RD performances are consistently and considerably better than the JPEG 2000 RD performance, for all LLF images and bitrates, with average BD-rate savings of 96.3% and 97.3%, respectively. This behavior shows the JPEG 2000 difficulty in exploiting redundancy significantly different from the one present in natural images, such as this spatial-angular redundancy.
- **LLFIC solutions vs. Non-royalty-free coding anchors:** The MuLE and PMV RD performances are consistently and considerably better than the HEVC Intra and VVC Intra RD performances, for all LLF images and bitrates, with average BD-rate savings of 93.6% and 91.1%, respectively. This is an expected behavior since the available angular redundancy is not exploited, or is barely exploited at the PIs borders, by both the HEVC Intra and VVC Intra anchors. Regarding the HEVC Inter anchor, both MuLE and PMV always present a better RD performance for all LLF images and bitrates, with average BD-rate savings of 25.5% and 49.5% respectively. This behavior shows that only by using 2D inter-view prediction (as in MuLE and PMV codecs) the available angular redundancy can be fully exploited; the HEVC Inter codec also exploits the angular redundancy but only between 'temporally' adjacent views (1D inter-view prediction). The PSNR gains with respect to the anchors tend to reduce as the bitrate increases, for all LLF images. At low bitrates, all data is 'difficult' to code; however, if the coding solution is appropriately 'equipped' with tools targeting the LLF spatial-angular redundancy exploitation, such as 2D inter-view prediction, the coding efficiency can be greatly improved; this is precisely what happens with both MuLE and PMV when compared to the anchor coding solutions, notably HEVC Intra/Inter and VVC Intra.
- **MuLE vs. PMV:** The PMV RD performance is consistently better than the MuLE RD performance for low to medium bitrates, for all LLF images. However, at the highest bitrate, the MuLE RD performance gets similar to or better than the PMV RD performance. This behavior indicates that, as the bitrate and quality increase, the spatial-angular redundancy is more efficiently exploited through a 4D transform than using inter-view prediction; this may be explained by the difficulty in obtaining a good enough inter-view prediction (resulting in a low residue to code) as the bitrate increases.

Figure 13 shows the SSIM-Y RD performance for the best performing royalty-free and non-royalty-free LLFIC solutions, namely MuLE and PMV, and all the LLFIC anchor coding solutions presented in Section V.C. From these charts, the following conclusions may be draw:

- **LLFIC solutions vs. Royalty-free coding anchor:** The SSIM RD performances for both MuLE and PMV are consistently and considerably better than the JPEG 2000 SSIM RD performance, for all LLF images and bitrates. This behavior shows that the 4D redundancy exploitation impacts on the perceptual quality of the decoded PIs; although both the LLFIC solutions and the JPEG anchor exploit the available spatial-angular redundancy, JPEG 2000 exploits it in a less efficient way (for the reason presented in Section V.D.1), thus resulting into a lower perceptual (decoded) quality.
- **LLFIC solutions vs. Non-royalty-free coding anchors:** The SSIM RD performances for both MuLE and PMV are also consistently and considerably better than the HEVC Intra and VVC Intra SSIM RD performances, for all LLF images and bitrates. Regarding the HEVC Inter anchor, both MuLE and PMV overcome the SSIM RD performance, for all LLF images and bitrates, although by a rather small margin at the highest bitrate. This behavior indicates that, at the highest bitrate, the LLFIC decoded qualities are so high that, perceptually speaking, they all look alike.
- **MuLE vs. PMV:** The PMV SSIM RD performance is consistently better than the MuLE SSIM RD performance, for all LLF images and bitrates except for the highest bitrate, where both coding solutions achieve a similar performance (for the reason above mentioned). This behavior may be explained by the fact that the PMV performance, which is based on the royalties-burdened MV-HEVC standard, benefits from the usage of powerful coding tools, e.g. CABAC based entropy coding, which the MuLE codec avoids to stay royalty-free. Moreover, while PMV uses very mature tools, resulting from decades of research, MuLE uses some recently designed tools, still without much refinement and optimization.

In summary, some main conclusions may be drawn from the previous comparative RD performance assessment:

- The PI representation basis seems to be better performing than the MI basis when it comes to 4D redundancy exploitation; moreover, no metadata is required at the decoder side for rendering/display purposes.
- The data type coded for the best performing LLFIC solutions is texture-only, so not requiring any geometry-related data; in fact, estimating high quality geometry-related data, e.g. depth or disparity, is a rather complex task and it is only worth to exploit this type of predictions if good estimations are available, what may be critically difficult for certain types of content.
- Adopting a data structure which is intrinsically 4D seems to allow a more natural and efficient exploitation of the spatial-angular redundancy, which is naturally 4D.

Among the best performing LLFIC solutions there is one that is royalty-free and has recently been adopted by the JPEG Pleno standard. Considering that the image coding landscape is typically royalty-free, this feature might be paramount for its market adoption in opposition to other efficient LLFIC solutions which are, however, royalties burdened. This will, naturally, depend on the application scenario requirements and proposed licensing model.

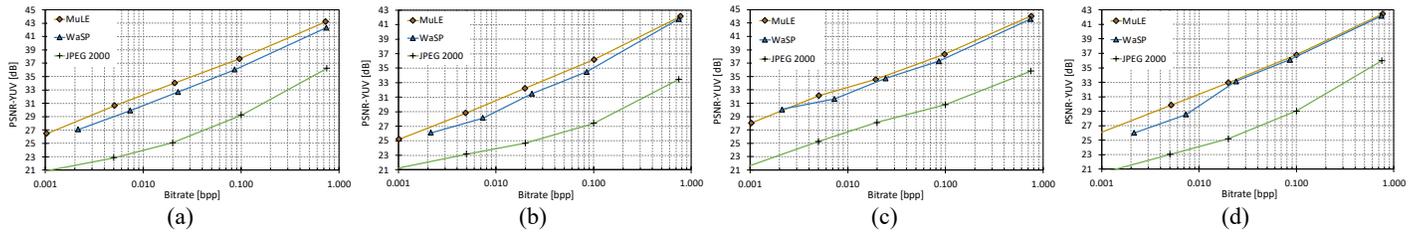

Figure 10 – PSNR-YUV RD performance for royalty-free LLF coding solutions: (a) *Bikes*; (b) *Danger de Mort*; (c) *Stone Pillars Outside*; and (d) *Fountain&Vincent2*.

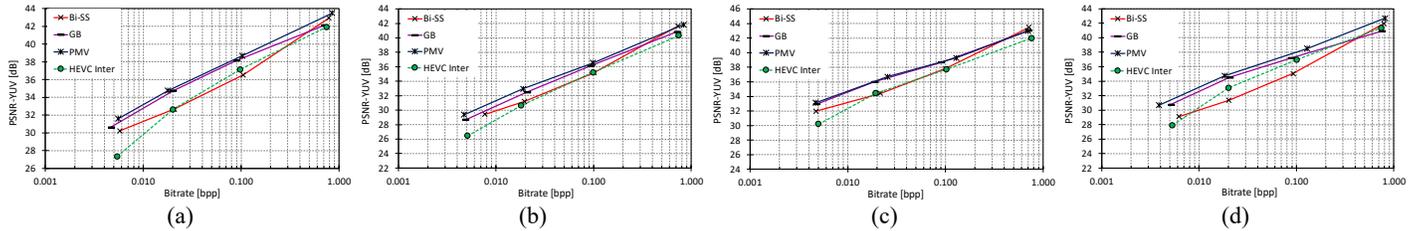

Figure 11 – PSNR-YUV RD performance for non-royalty-free LLF coding solutions: (a) *Bikes*; (b) *Danger de Mort*; (c) *Stone Pillars Outside*; and (d) *Fountain&Vincent2*.

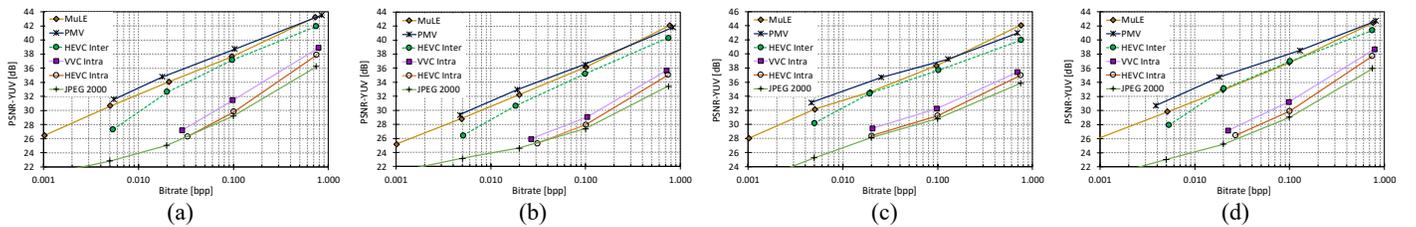

Figure 12 – PSNR-YUV RD performance for the best performing royalty-free and non-royalty-free LLF coding solutions: (a) *Bikes*; (b) *Danger de Mort*; (c) *Stone Pillars Outside*; and (d) *Fountain&Vincent2*.

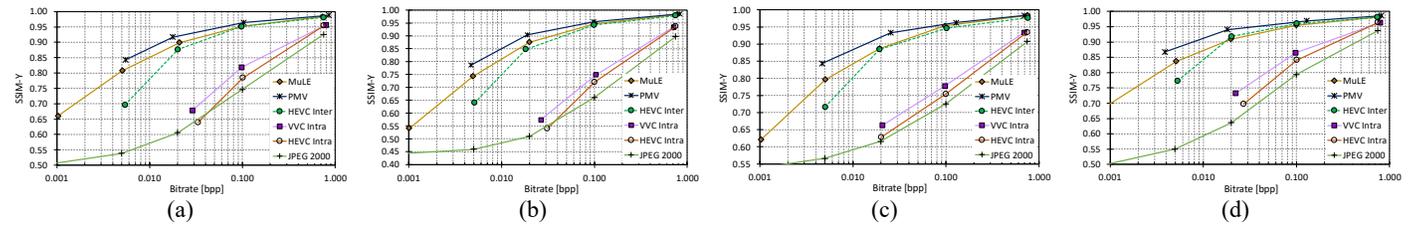

Figure 13 – SSIM-Y RD performance for the best performing royalty-free and non-royalty-free LLF coding solutions: (a) *Bikes*; (b) *Danger de Mort*; (c) *Stone Pillars Outside*; and (d) *Fountain&Vincent2*.

ACKNOWLEDGMENTS

The authors would like to thank C. Conti, E. A. B. da Silva, I. Viola, M. Pereira, P. Astola, and W. Ahmad for kindly and promptly providing performance results for their methods under the JPEG Pleno Light Field Coding Common Test Conditions.

REFERENCES

[1] E. H. Adelson and J. R. Bergen, "The plenoptic function and the elements of early vision", The MIT Press, Cambridge, Mass., 1991.
[2] S. J. Gortler *et al.*, "The lumigraph", in *ACM SIGGRAPH'96*, New Orleans, LA, USA, Aug. 1996.
[3] JPEG Convenor, "JPEG Pleno abstract and executive summary", Doc. ISO/IEC JTC 1/SC 29/WG1 N6922, Sydney, Australia, Feb. 2015.
[4] JPEG Convenor, "JPEG Pleno call for proposals on light field coding", Doc. ISO/IEC JTC 1/SC 29/WG1 N74014, Geneva, Switzerland, Jan. 2017.
[5] *Information technology — JPEG Pleno Plenoptic image coding system — Part 2: Light Field Coding*, ISO/IEC CD 21794-2, ISO/IEC JTC 1/SC 29.
[6] F. Pereira, E. A. B. da Silva, and G. Lafruit, "Plenoptic imaging: representation and processing," *Academic Press Library in Signal Processing*, vol. 6, R. Chellappa and S. Theodoridis, Eds. Academic Press, pp. 75–11, 2018.
[7] https://en.wikipedia.org/wiki/Lytro
[8] https://raytrix.de/
[9] R. Ng, M. Levoy, M. Brédif, G. Duval, M. Horowitz and P. Hanrahan, "Light field photography with a hand-held plenoptic camera," Tech. Rep. CSTR 2005-02, Stanford University, Stanford, CA, USA, Apr. 2005.
[10] F. Pereira *et al.*, "JPEG Pleno light field coding common test conditions v3.2," Doc. ISO/IEC JTC 1/SC 29/WG1 N83029, Geneva, Switzerland, Mar. 2019.
[11] T. Georgiev and A. Lumsdaine, "Focused plenoptic camera and rendering," *Journal of Electronic Imaging*, vol. 19, no. 2, pp. 021106-1–021106-11, Apr–Jun 2010.
[12] C. Conti, J. Lino, P. Nunes, L. D. Soares, and P. L. Correia, "Improved spatial prediction for 3D holoscopic image and video coding," in *Eur. Signal Process. Conf.*, Barcelona, Spain, Aug. 2011, pp. 378–382.
[13] C. Conti, J. Lino, P. Nunes, L. D. Soares, and P. L. Correia, "Spatial prediction based on self-similarity compensation for 3D holoscopic image and video coding," in *IEEE Int. Conf. Image Process.*, Brussels, Belgium, Sept. 2011, pp. 961–964.
[14] C. Conti, P. Nunes, and L. D. Soares, "New HEVC prediction modes for 3D holoscopic video coding," in *IEEE Int. Conf. Image Process.*, Orlando, FL, USA, Sept. 2012, pp. 1325–1328.
[15] C. Conti, P. Nunes, and L. D. Soares, "Inter-layer prediction scheme for scalable 3-D holoscopic video coding," *IEEE Signal Process. Lett.*, vol. 20, no. 8, pp. 819–822, Aug. 2013.
[16] C. Conti, P. Nunes, and L. D. Soares, "Using self-similarity compensation for improving inter-layer prediction in scalable 3D holoscopic video coding," in *SPIE Optical Engeneering + Applications*, San Diego, CA, USA, Aug. 2013.
[17] Y. Li, M. Sjostrom, R. Olsson, and U. Jennehag, "Efficient intra prediction scheme for light field image compression," in *IEEE Int. Conf. Acoust., Speech, Signal Process.*, Florence, Italy, May 2014, pp. 539–543.
[18] L. F. R. Lucas *et al.*, "Locally linear embedding-based prediction for 3D holoscopic image coding using HEVC," in *Eur. Signal Process. Conf.*, Lisbon, Portugal, Sept. 2014, pp. 11–15.
[19] C. Perra, "On the coding of plenoptic raw images," in *Telecommunications*

*Forum*, Belgrade, Serbia, Nov. 2014, pp. 850–853.

[20] C. Choudhury and S. Chaudhuri, "Disparity based compression technique for focused plenoptic images," in *Indian Conf. Computer Vision Graphics and Image Processing*, Bangalore, India, Dec. 2014.

[21] C. Perra, "Lossless plenoptic image compression using adaptive block differential prediction," in *IEEE Int. Conf. Acoust., Speech, Signal Process.*, Brisbane, QLD, Australia, Apr. 2015, pp. 1231–1234.

[22] Y. Li, M. Sjostrom, and R. Olsson, "Coding of plenoptic images by using a sparse set and disparities," in *IEEE Int. Conf. Multimedia and Expo*, Turin, Italy, Jun. 2015.

[23] D. Liu, P. An, R. Ma, and L. Shen, "Disparity compensation based 3D holoscopic image coding using HEVC," in *IEEE China Summit and Int. Conf. Signal Information Process.*, Chengdu, China, Jul. 2015, pp. 201–205.

[24] A. Dricot, J. Jung, M. Cagnazzo, B. Pesquet, and F. Dufaux, "Integral images compression scheme based on view extraction," in *Eur. Signal Process. Conf.*, Nice, France, Aug. 2015, pp. 101–105.

[25] F. Dai, J. Zhang, Y. Ma, and Y. Zhang, "Lenselet image compression scheme based on subaperture images streaming," in *IEEE Int. Conf. Image Process.*, Quebec City, QC, Canada, Sept. 2015, pp. 4733–4737.

[26] Y. Li, M. Sjöström, R. Olsson, and U. Jennehag, "Scalable coding of plenoptic images by using a sparse set and disparities," *IEEE Trans. Image Process.*, vol. 25, no. 1, pp. 80–91, Jan. 2016.

[27] C. Conti, L. D. Soares, and P. Nunes, "HEVC-based 3D holoscopic video coding using self-similarity compensated prediction," *Signal Process.: Image Commun.*, vol. 42, pp. 59–78, Mar. 2016.

[28] C. Conti, P. Nunes, and L. D. Soares, "HEVC-based light field image coding with bi-predicted self-similarity compensation," in *IEEE Int. Conf. Multimedia and Expo Workshops*, Seattle, WA, USA, Jul. 2016.

[29] Y. Li, R. Olsson, and M. Sjostrom, "Compression of unfocused plenoptic images using a displacement intra prediction," in *IEEE Int. Conf. Multimedia and Expo Workshops*, Seattle, WA, USA, Jul. 2016.

[30] D. Liu, L. Wang, L. Li, X. Zhiwei, F. Wu, and Z. Wenjun, "Pseudo-sequence-based light field image compression," in *IEEE Int. Conf. Multimedia and Expo Workshops*, Seattle, WA, USA, Jul. 2016.

[31] R. Monteiro *et al.*, "Light field HEVC-based image coding using locally linear embedding and self-similarity compensated prediction," in *IEEE Int. Conf. Multimedia and Expo Workshops*, Seattle, WA, USA, Jul. 2016.

[32] C. Perra and P. A. A. Assuncao, "High efficiency coding of light field images based on tiling and pseudo-temporal data arrangement," in *IEEE Int. Conf. Multimedia and Expo Workshops*, Seattle, WA, USA, Jul. 2016.

[33] P. Helin, P. Astola, B. Rao, and I. Tabus, "Sparse modelling and predictive coding of subaperture images for lossless plenoptic image compression," in *3DTV-Conference*, Hamburg, Germany, Jul. 2016.

[34] Y. Li, M. Sjöström, R. Olsson, and U. Jennehag, "Coding of focused plenoptic contents by displacement intra prediction," *IEEE Trans. Circuits Syst. Video Technol.*, vol. 26, no. 7, pp. 1308–1319, Jul. 2016.

[35] D. Liu, P. An, R. Ma, C. Yang, L. Shen, and K. Li, "Three-dimensional holoscopic image coding scheme using high-efficiency video coding with kernel-based minimum mean-square-error estimation," *J. Electron. Imaging*, vol. 25, no. 4, pp. 1–9, Jul. 2016.

[36] D. Liu, P. An, R. Ma, C. Yang, and L. Shen, "3D holoscopic image coding scheme using HEVC with Gaussian process regression," *Signal Process.: Image Commun.*, vol. 47, pp. 438–451, Sept. 2016.

[37] A. Dricot, J. Jung, M. Cagnazzo, B. Pesquet, and F. Dufaux, "Improved integral images compression based on multi-view extraction," in *SPIE Optical Engineering + Applications*, San Diego, CA, USA, Sept. 2016.

[38] C. Perra, "Light field image compression based on preprocessing and high efficiency coding," in *Telecommunications Forum*, Belgrade, Serbia, Nov. 2016.

[39] S. Zhao, Z. Chen, K. Yang, and H. Huang, "Light field image coding with hybrid scan order," in *Visual Communications and Image Processing*, Chengdu, China, Nov. 2016.

[40] L. Yang, P. An, D. Liu, and R. Ma, "3D holoscopic images coding scheme based on viewpoint image rendering," *Int. Forum Digital TV and Wireless Multimedia Communication*, Shanghai, China, Nov. 2016, pp. 318–327.

[41] D. Liu, P. An, T. Du, R. Ma, and L. Shen, "An improved 3D holoscopic image coding scheme using HEVC based on Gaussian mixture models," *Int. Forum Digital TV and Wireless Multimedia Communication*, Shanghai, China, Nov. 2016, pp. 276–285.

[42] R. Zhong, S. Wang, B. Cornelis, Y. Zheng, J. Yuan, and A. Munteanu, "L1-optimized linear prediction for light field image compression," in *Picture Coding Symposium*, Nuremberg, Germany, Dec. 2016.

[43] H. Han, X. Jin, and Q. Dai, "Plenoptic image compression based on linear transformation and interpolation," in *Asia-Pacific Signal and Information Process. Association Annual Summit and Conf.*, Jeju, South Korea, Dec. 2016.

[44] C. Perra and D. Giusto, "JPEG 2000 compression of unfocused light field images based on lenslet array slicing," in *IEEE Int. Conf. Consumer Electronics*, Las Vegas, NV, USA, Jan. 2017.

[45] D. Liu, P. An, C. Yang, R. Ma, and L. Shen, "Coding of 3D holoscopic image by using spatial correlation of rendered view images," *IEEE Int. Conf. Acoust., Speech, Signal Process.*, New Orleans, LA, USA, Mar. 2017, pp. 2002–2006.

[46] X. Jiang, M. Le Pendu, R. A. Farrugia, S. S. Hemami, and C. Guillemot, "Homography-based low rank approximation of light fields for compression," in *IEEE Int. Conf. Acoust., Speech, Signal Process.*, New Orleans, LA, USA, Mar. 2017, pp. 1313–1317.

[47] L. Li, Z. Li, B. Li, D. Liu, and H. Li, "Pseudo-sequence-based 2-D hierarchical coding structure for light-field image compression," in *Data Compression Conference*, Snowbird, UT, USA, Apr. 2017, pp. 131–140.

[48] H. P. Hariharan, T. Lange, and T. Herfet, "Low complexity light field compression based on pseudo-temporal circular sequencing," in *IEEE Int. Symp. Broadband Multimedia Systems and Broadcasting*, Cagliari, Italy, Jun.2017.

[49] C. Perra and D. Giusto, "Raw light field image compression of sliced lenslet array," in *IEEE Int. Symp. Broadband Multimedia Systems and Broadcasting*, Cagliari, Italy, Jun.2017.

[50] I. Schiopu, M. Gabbouj, A. Gotchev, and M. M. Hannuksela, "Lossless compression of subaperture images using context modeling," *3DTV-Conference*, Copenhagen, Denmark, Jun. 2017.

[51] R. Verhack, T. Sikora, L. Lange, R. Jongebloed, G. Van Wallendael, and P. Lambert, "Steered mixture-of-experts for light field coding, depth estimation, and processing," in *IEEE Int. Conf. Multimedia and Expo*, Hong Kong, China, Jul. 2017, pp. 1183–1188.

[52] H. Han, X. Jin, and Q. Dai, "Lenslet image compression based on image reshaping and macro-pixel Intra prediction," in *IEEE Int. Conf. Multimedia and Expo*, Hong Kong, China, Jul. 2017, pp. 1177–1182.

[53] X. Jiang, M. Le Pendu, and C. Guillemot, "Light field compression using depth image based view synthesis," in *IEEE Int. Conf. Multimedia and Expo Workshops*, Hong Kong, China, Jul. 2017, pp. 19–24.

[54] C. Conti, L. D. Soares, and P. Nunes, "Weighted bi-prediction for light field image coding," in *SPIE Optical Engineering + Applications*, San Diego, CA, USA, Aug. 2017, no. 351.

[55] I. Tabus and P. Helin, "Microlens image sparse modelling for lossless compression of plenoptic camera sensor images," in *Eur. Signal Process. Conf.*, Kos, Greece, Aug. 2017, pp. 1907–1911.

[56] W. Ahmad, R. Olsson, and M. Sjoestroem, "Interpreting plenoptic images as multi-view sequences for improved compression," in *IEEE Int. Conf. Image Process.*, Beijing, China, Sept. 2017, pp. 4557–4561.

[57] S. Zhao and Z. Chen, "Light field image coding via linear approximation prior," in *IEEE Int. Conf. Image Process.*, Beijing, China, Sept. 2017, pp. 4562–4566.

[58] I. Tabus, P. Helin, and P. Astola, "Lossy compression of lenslet images from plenoptic cameras combining sparse predictive coding and JPEG 2000," in *IEEE Int. Conf. Image Process.*, Beijing, China, Sept. 2017, pp. 4567–4571.

[59] R. Zhong, S. Wang, B. Cornelis, Y. Zheng, J. Yuan, and A. Munteanu, "Efficient directional and L1-optimized intra-prediction for light field image compression," in *IEEE Int. Conf. Image Process.*, Beijing, China, Sept. 2017, pp. 1172–1176.

[60] Y.-H. Chao, G. Cheung, and A. Ortega, "Pre-demosaic light field image compression using graph lifting transform," in *IEEE Int. Conf. Image Process.*, Beijing, China, Sept. 2017, pp. 3240–3244.

[61] H. Han, X. Jin, and Q. Dai, "Lenslet image compression using adaptive macropixel prediction," in *IEEE Int. Conf. Image Process.*, Beijing, China, Sept. 2017, pp. 4008–4012.

[62] C. Jia *et al.*, "Optimized inter-view prediction based light field image compression with adaptive reconstruction," in *IEEE Int. Conf. Image Process.*, Beijing, China, Sept. 2017, pp. 4572–4576.

[63] C. Jia, Y. Yang, and X. Zhang, "Light field image compression with sub-apertures reordering and adaptive reconstruction," in *Pacific-Rim Conference on Multimedia*, Harbin, China, Sept. 2017.

[64] P. Helin, P. Astola, B. Rao, and I. Tabus, "Minimum description length sparse modeling and region merging for lossless plenoptic image compression," *IEEE J. Sel. Topics Signal Process.*, vol. 11, no. 7, pp. 1146–1161, Oct. 2017.

[65] X. Jiang, M. Le Pendu, R. A. Farrugia, and C. Guillemot, "Light field compression with homography-based low-rank approximation," *IEEE J. Sel. Topics Signal Process.*, vol. 11, no. 7, pp. 1132–1145, Oct. 2017.

[66] X. Jin, H. Han, and Q. Dai, "Image reshaping for efficient compression of plenoptic content," *IEEE J. Sel. Topics Signal Process.*, vol. 11, no. 7, pp. 1173–1186, Oct. 2017.

[67] L. Li, Z. Li, B. Li, D. Liu, and H. Li, "Pseudo-sequence-based 2-D hierarchical coding structure for light-field image compression," *IEEE J. Sel. Topics Signal Process.*, vol. 11, no. 7, pp. 1107–1119, Oct. 2017.

[68] R. Monteiro, P. Nunes, N. Rodrigues, and S. M. M. Faria, "Light field image coding using high-order intrablock prediction," *IEEE J. Sel. Topics Signal Process.*, vol. 11, no. 7, pp. 1120–1131, Oct. 2017.

[69] J. M. Santos, P. A. A. Assuncao, L. A. da Silva Cruz, L. M. N. Távora, R. Fonseca-Pinto, and S. M. M. Faria, "Lossless light-field compression using reversible colour transformations," in *Int. Conf. Image Processing Theory, Tools and Applications*, Montreal, QC, Canada, Nov. 2017.

[70] W. Zhang, D. Liu, Z. Xiong, and J. Xu, "SIFT-based adaptive prediction structure for light field compression," in *Visual Communications and Image Processing*, St. Petersburg, FL, USA, Dec. 2017.

[71] C. Perra, "Light field coding based on flexible view ordering for unfocused plenoptic camera images," *Int. Journal of Applied Engineering Research*, vol. 12, no. 21, pp. 10563–10569, 2017.

[72] J. Chen, J. Hou, and L.-P. Chau, "Light field compression with disparity-guided sparse coding based on structural key views," *IEEE Trans. Image Process.*, vol. 27, no. 1, pp. 314–324, Jan. 2018.

[73] D. Liu, P. An, R. Ma, C. Yang, L. Shen, and K. Li, "Scalable coding of 3D holoscopic image by using a sparse interlaced view image set and disparity map," *Multimed. Tools Appl.*, vol. 77, no. 1, pp. 1261–1283, Jan. 2018.

[74] A. Miyazawa, Y. Kameda, T. Ishikawa, I. Matsuda, and S. Itoh, "Lossless coding of light field camera data captured with a micro-lens array and a color filter," in *Int. Workshop on Advanced Image Technology*, Chiang Mai, Thailand, Jan. 2018.

[75] Y. Y. Liu, C. Zhu, and M. Mao, "Light field image compression based on quality aware pseudo-temporal sequence," *Electron. Lett.*, vol. 54, no. 8, pp. 500–501, Feb. 2018.

[76] C. Conti, P. Nunes, and L. D. Soares, "Light field image coding with jointly estimated self-similarity bi-prediction," *Signal Process.: Image Commun.*, vol. 60, pp. 144–159, Feb. 2018.

[77] B. Guo, Y. Han, and J. Wen, "Convex optimization based bit allocation for light field compression under weighting and consistency constraints," in *Data Compression Conference*, Snowbird, UT, USA, Mar. 2018, pp. 107–116.

[78] M. Rizkallah, X. Su, T. Maugey, and C. Guillemot, "Graph-based transforms for predictive light field compression based on super-pixels," in *IEEE Int. Conf. Acoust., Speech, Signal Process.*, Calgary, AB, Canada, Apr. 2018, pp. 1718–1722.

[79] J. M. Santos, P. A. A. Assuncao, L. A. da Silva Cruz, L. M. N. Tavora, R. Fonseca-Pinto, and S. M. M. Faria, "Lossless coding of light field images based on minimum-rate predictors," *Journal of Visual Communication and Image Representation*, vol. 54, pp. 21–30, Jul. 2018.

[80] Z. You, P. An, and D. Liu, "Scalable kernel-based minimum mean square error estimate for light-field image compression," *EURASIP Journal on Image and Video Processing*, vol. 2018, Jul. 2018.

[81] X. Huang, P. An, L. Shen, and R. Ma, "Efficient light field images compression method based on depth estimation and optimization," *IEEE Access*, vol. 6, pp. 48984–48993, Aug. 2018.

[82] X. Jin, H. Han, and Q. Dai, "Plenoptic image coding using macropixel-based Intra prediction," *IEEE Trans. Image Process.*, vol. 27, no. 8, pp. 3954–3968, Aug. 2018.

[83] B. Guo, J. Wen, and Y. Han, "Two-pass light field image compression for spatial quality and angular consistency," *ArXiv*, no. 61521002, pp. 1–17, Aug. 2018.

[84] J. Garrote, C. Brites, J. Ascenso, and F. Pereira, "Lenslet light field imaging scalable coding," in *Eur. Signal Process. Conf.*, Rome, Italy, Sept. 2018.

[85] R. Monteiro, P. Nunes, S. M. M. Faria, and N. Rodrigues, "Light field image coding using high order prediction training," in *Eur. Signal Process. Conf.*, Rome, Italy, Sept. 2018, pp. 1845–1849.

[86] X. Su, M. Rizkallah, T. Maugey, and C. Guillemot, "Rate-distortion optimized super-ray merging for light field compression," in *Eur. Signal Process. Conf.*, Rome, Italy, Sept. 2018, pp. 1850–1854.

[87] I. Viola, H. Petric, P. Frossard, and T. Ebrahimi, "A graph learning approach for light field image compression," in *SPIE Optical Engineering + Applications*, San Diego, CA, USA, Sept. 2018.

[88] H. Han, J. Xin, and Q. Dai, "Plenoptic image compression via simplified subaperture projection," in *Pacific Rim Conference on Multimedia*, Hefei, China, Sept. 2018, pp. 274–284.

[89] M. B. de Carvalho *et al.*, "A 4D DCT-based lenslet light field codec," in *IEEE Int. Conf. Image Process.*, Athens, Greece, Oct. 2018, pp. 435–439.

[90] W. Ahmad, R. Olsson, and M. Sjostrom, "Towards a generic compression solution for densely and sparsely sampled light field data," in *IEEE Int. Conf. Image Process.*, Athens, Greece, Oct. 2018, pp. 654–658.

[91] N. Bakir, W. Hamidouche, and O. Déforges, "Light field image compression based on convolutional neural networks and linear approximation," in *IEEE Int. Conf. Image Process.*, Athens, Greece, Oct. 2018, pp. 1128–1132.

[92] R. Conceição, M. Porto, B. Zatt, and L. Agostini, "LF-CAE: Context-adaptive encoding for lenslet light fields using HEVC," in *IEEE Int. Conf. Image Process.*, Athens, Greece, Oct. 2018, pp. 3174–3178.

[93] I. Schiopu and A. Munteanu, "Macro-pixel prediction based on convolutional neural networks for lossless compression of light field images," in *IEEE Int. Conf. Image Process.*, Athens, Greece, Oct. 2018, pp. 445–449.

[94] P. Astola and I. Tabus, "WaSP: Hierarchical warping, merging, and sparse prediction for light field image compression," in *European Workshop on Visual Information Processing*, Tampere, Finland, Nov. 2018.

[95] C. Conti, L. D. Soares, and P. Nunes, "Light field coding with field-of-view scalability and exemplar-based interlayer prediction," *IEEE Trans. Multimedia*, vol. 20, no. 11, pp. 2905–2916, Nov. 2018.

[96] D. Liu, P. An, R. Ma, and L. Shen, "Hybrid linear weighted prediction and intra block copy based light field image coding," *Multimed. Tools Appl.*, vol. 77, no. 24, pp. 31929–31951, Dec. 2018.

[97] J. Hou, J. Chen, and L. P. Chau, "Light field image compression based on bi-level view compensation with rate-distortion optimization," *IEEE Trans. Circuits Syst. Video Technol.*. vol. 29, no. 2, pp. 517–530, Feb. 2019.

[98] R. Zhong, I. Schiopu, B. Cornelis, S. P. Lu, J. Yuan, and A. Munteanu, "Dictionary learning-based, directional and optimized prediction for lenslet image coding," *IEEE Trans. Circuits Syst. Video Technol.*. vol. 29, no. 4, pp. 1116–1129, Apr 2019.

[99] "Verification Model Software Version 2.0 on JPEG Pleno Light Field Coding," Doc. ISO/IEC JTC1/SC29/WG1 N82046, Lisbon, Portugal, Jan. 2019.

[100] https://github.com/braites/LLFIC

[101] ITU-T Q.6/SG 16 and ISO/IEC JTC 1/SC 29/WG 11, "High Efficiency Video Coding (HEVC) reference software HM." [Online]. Available: https://hevc.hhi.fraunhofer.de/trac/hevc/browser/trunk.

[102] D. S. Taubman and M. W. Marcellin, *JPEG 2000: Image Compression Fundamentals, Standards and Practices*. Norwell, MA, USA: Kluwer, 2002.

[103] "Kakadu software," [Online]. Available: http://kakadusoftware.com.

[104] T. C. Wang, A. A. Efros, and R. Ramamoorthi, "Occlusion-aware depth estimation using light-field cameras," in *IEEE Int. Conf. Computer Vision*, Santiago, Chile, Dec 2015, pp. 3487–3495.

[105] "JPEG Pleno database," [Online]. Available: https://jpeg.org/jpegpleno/plenodb.html.

[106] Z. Wang, A. C. Bovik, H. R. Sheikh, and E. P. Simoncelli, "Image quality assessment: from error visibility to structural similarity", *IEEE Trans. Image Process.*, vol. 13, no. 4, pp. 600–612, Apr. 2004.

[107] *http://www.mathworks.com/matlabcentral/fileexchange/27798-bjontegaardmetric/content/bjontegaard.m*

[108] "OpenJPEG," [Online]. Available: http://www.openjpeg.org/.

[109] G. J. Sullivan *et al.*, "Overview of the high efficiency video coding (HEVC) standard", *IEEE Trans. Circuits Syst. Video Technol.*, vol. 22, no. 12, pp. 1649–1668, Dec. 2012.

[110] https://bitbucket.org/multicoreware/x265/

[111] "Versatile video coding," [Online]. Available: https://mpeg.chiariglione.org/standards/mpeg-i/versatile-video-coding

[112] "VVC VTM reference software," [Online]. Available: https://vcgit.hhi.fraunhofer.de/jvet/VVCSoftware_VTM

[113] M. P. Pereira *et al.*, "A geometric space-view redundancy descriptor for light fields: Predicting the compression potential of the JPEG Pleno light field datasets," *IEEE Int.Workshop Multimedia Signal Processing*, Luton, UK, Oct. 2017, pp. 1-6.

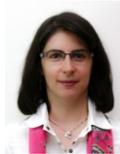

**Catarina Brites** (M'11) received the E.E., M.Sc., and Ph.D. degrees in electrical and computer engineering from the Instituto Superior Técnico (IST), Universidade Técnica de Lisboa, Lisbon, Portugal, in 2003, 2005, and 2011, respectively. She is currently a Postdoctoral Researcher with the Multimedia Signal Processing Group, Instituto de Telecomunicações, IST.

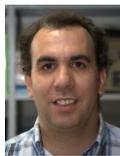

**João Ascenso** (SM'18) received the E.E., M.Sc., and Ph.D. degrees from Instituto Superior Técnico (IST), Universidade Técnica de Lisboa, Lisbon, Portugal, in 1999, 2003, and 2010, respectively, all in electrical and computer engineering. He is currently an Assistant Professor with the Department of Electrical and Computer Engineering, IST, and member of Instituto de Telecomunicações (IT).

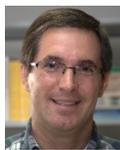

**Fernando Pereira** (F'08) received the B.S., M.Sc., and Ph.D. degrees in electrical and computer engineering from the Instituto Superior Técnico (IST), Universidade de Lisboa, Lisbon, Portugal, in 1985, 1988, and 1991, respectively. He is currently with the Electrical and Computer Engineering Department, IST, and Instituto de Telecomunicações.